\documentstyle[amssymb]{article}

\def\al{\alpha}
\def\ga{\gamma}
\def\om{\omega}

\def\iy{\infty}
\def\De{\Delta}

\def\si{\sigma}

\def\dac{\displaystyle\frac}

\def\dil{\displaystyle\int\limits}

\begin{document}

\begin{center}

\vspace{1cm}

{\large{\bf

Compyter analysis of undulators with block-periodic stucture

\vspace{1cm}

A.F. Medvedev, M.L. Schinkeev}}

\medskip

\it{Tomsk Polytechnic University}

\end{center}

\vspace{2cm}

{\bf Abstract}

\medskip

Methods to detect the spectral sensitivity of an object using undulator
radiation  without monochromators  or any  other spectral devices are
developed. The  spectral transmission  function of  the object  is
calculated  from its  response to  the spectrum-integral undulator
radiation with the known spectral distribution. This response is measured
as a function of the electron energy.

\medskip

{\bf l. Introduction}

\medskip

At  present,  synchrotron  radiation  (SR)   is  widely
used in  spectroscopy  as  a  standard  source,   the  monochromatic
components  usually   being  obtained   by  mono- chromators  or  other
spectral  devices.  However,  a  large intensity  loss  and variation  of
the  transmission function during  the  operating  time  are  inherent to
such measurements.

Recently,  undulator  radiation  (UR)  has   been
discussed  as  an  alternative.  As  is  known  [1], monochroma-
tization  of  UR  can  be  partially  achieved  by increasing
the  period  number.  UR  serving  as  a  standard  source in
spectroscopy  without    monochromators   has    been   discussed  in ref.
[2]. It  appears that  the UR  resolution is not  only  limited by  the
spectral  line width  that depends on  the  period  number, but  is also
limited by  the spread of  angles  and  electron  energies,  the  undulator
magnetic field  nonuniformity  over  the   beam  cross   section,  the
finite diaphragm size, and other factors.

In  this   connection,  a   monochromatorless  computer
spectroscopy   method   (MCS   method)   has   been  proposed
[3,4],  the  computer  algorithm  playing  the  part  of  the
monochromator.  The  point  is  that  the  radiation  from an
undulator  installed  in  the  synchrotron  ring is  not pure
UR  as  assume'd in  the ideal  theoretical model  [1,5], but
also   includes  the   SR  components   from  the   edges  of
bending  magnets  and  focusing  elements  adjacent   to  the
undulator.  In  practice,  one  uses the  frequency partition
of  SR  and  UR  spectra  to  exclude  the  admixed   SR  and
to  conserve  the  ideal  UR  properties  [6]. But  we cannot
admit  this  method  to  be  consistent  with   the  requirements  of
metrology,  according  to which  all the  ideal UR properties    (angular
monochromatization,   polarization, independence  of  the  UR  spectral
form  from  the particle energy)  need  to  be  confirmed  by   adequate
quantitative measurements.

\medskip

{\bf 2. Amplitude-time modulation}

\medskip

In the MCS method, the electron-energy-invariant
spectral form of the. radiation source turns out to be a
kernel of the integral equation, the solution of which
being just the MCS problem. In order to make the UR
kernel  metrologically pure, we have to exclude the admixed SR. To this end
we can use a procedure consistng of a series of consequent measurements
of the undulator radiation at various states of the undulator magnetic
system and the subsequent combination of these results. Such a procedure
will be referred to as the amplitude - time modulation (ATM).

Parameters varying in the ATM process are those of the undulator, which do
not disturb the properties of the admixed SR. This means that the phase
relations for the SR components must be invariant during ATM.  This implies
the constancy of the time $t$ for a charge travelling along a straight
section of length $l$ between the edge elements adjacent to the undulator:
\begin{equation}
t=\dac{1}{c}\Big[l+\dac{l+k^2L}{2\ga^2}\Big].
\end{equation}
Here  $L$  is  the  undulator   length,  $\ga$  the   Lorenz  factor
of  the  charge  ($\ga^{-1}\ll1$),  $c$ is  the light  velocity, and  $k$
is the undulator dipole parameter.  The  directional   modulation  of
the  magnetic   field  in the  undulator  or  in  some  of  its
blocks  is  a  special  case of  ATM  since  the  electron  transit
time  is   conserved.  The magnetic  field  in   any  block   satisfies
the   balancing  condition,   i.e.,   the   field  integral   over  the
block  length vanishes.

Suppose   $A(\om)$   and  $B(\om)$    are   the
complex   Fourier amplitudes   of   the   ideal   UR   and   the
SR,  respectively.  Every    $A(\om)$    corresponds    to    one
undulator    magnetic field  state.  Let  us  consider  the  following
four  states: the basic  state  $A$;  the  state  $(  -  A)$,  which
differs from  $A$ in that the undulator magnetic field is switched on
with an orientation opposite to the field in the elements forming the
electron  orbit  and  adjoining  the  undulator; the phase-discontinuous
state $\tilde{A}$ obtained by  the reverse-sign switching  of  the  magnetic
field in   some  undulator blocks; and finally $( - \tilde{A})$. Adding the
radiation intensi- ties in the first and second states and subtracting
those in the third and the fourth states, we obtain
\begin{equation}
|A+B|^2+|-A+B|^2-|\tilde{A}+B|^2-|-\tilde{A}+B|^2=2\big[|A|^2-|\tilde{A}
|^2\big].
\end{equation}
Thus, the four-step ATM with phase switching enables
us to exclude the SR and to obtain the metrologically
pure UR kernel for MCS as an element of the ideal UR
(eq. (2)). The practical realization will be especially
simple when one ATM step corresponds to one acceleration cycle. In this
case, we need only four cycles to achieve a metrologically pure procedure
with UR.

\medskip

{\bf 3. Undulator magnetic system}

\medskip

Among the various possible realizations of the undulator magnetic
system for MCS with UR, the electromagnetic ironless system is preferable
since only such a system has the desired properties, such as linearity,
predictability, reiteration and the ability to change undulator states
quickly. A block-periodic organization of the ironless electromagnetic
undulator system in which the resulting distribution of the magnetic field
is a superposition of fields from standard elements switching on mth given
weights, seems to be the most suitable one for many types of ATM. In the
general case, because of the common standard elements, the blocks may
overlap one another with the equivalent summarized overlap weight.

The MCS technique needs no monochromatization devices since the computer
algorithm plays the part of the monochromator. Thus we avoid the many order
waste of the source intensity, which is usually inevitable
with radiation monochromatization. It appears from
this that an undulator with the dipole regime of UR
excitation $(k\ll1)$ will be quite effective in achieving an
adequate reaction of the object in the MCS technique.
The dipole regime is favourable also with regard to the
'operation of the undulator magnetic system, since the
decreased heat and electromagnetic loads permit the
ironless variant of this system to be used.

Therefore, we can now regard the undulator dipole
regime as the basic one for the MCS technique with
UR, and so the model investigations of MCS with UR,
based on the dipole approximation expression for UR,
become legitimate.

In ATM it is important to provide the coincidence of
the low-frequency trend components of the motion in
the  undulator  for  different  steps  of  the  ATM  period.
The  amplitudes  of  the  low-frequency  components  of  the
spectrum  depend  essentially  on  the  values  of  the  integrals
\begin{center}
$J_1=\dil_0^b{H(x)dx}, ~~J_2=\dil_0^b{dx'}
\dil_0^{x'}{H(x)dx}, ~.  .  .$
\end{center}
where  $H(x)$ is  the  magnetic  field along  the undulator axis and $b$ is
the block length. We  shall call  the motion of a  particle  in  a  block
an  $m$-time balanced  motion, when $J_1 =  J_2 = ... = J_m  = 0$.  The
Fourier structure of the  UR  line  of   a block with  the   motion
balancing degree $m$ contains the factor)
\begin{center}
$|\cos^m{\bar\om}\sin{\bar\om(N-m)}/sin{\bar\om}|^2,$
\end{center}
where $\bar\om=\pi(\nu+1)/2, ~\nu=\eta(1+\psi^2)$  is the  number of
the UR harmonic  at  an angle $\theta$ to  the motion axis,
$\psi=\ga\theta/\sqrt{1+k^2}, \eta=p\om(1+k^2)/{2\pi c\ga^2}, \om$ is the
radiation frequency, $p$ is the magnetic field half-period length, and $N$
is the  number of standard elements  in the block.  The low-frequency
asymptote   for such a spectrum  is $\nu^{2m}$ .  It appears from this
that in the region $0 < \nu < 1$ the  UR spectral density at  a given
direction $\theta$ is suppressed  and the number  of spectral function zeros
decreases  when $m$ increases.  This causes  the oscillating part  of the
angle-integral UR spectrum to  be depressed.  As a result, the difference
UR  kernel of  the type  of eq.  (2)  formed  by the  ATM is localized in a
region  of the high-frequency cut-off   of the fundamental harmonic.  The
maximal balancing  degree in a  block consisting of $N$ standard elements
equals $N -  l$.  In this  case, the maximal  smoothness of  the UR
integral spectrum and the  maximal frequency,  angle  and polarization
localization of the difference kernel of type (2) are obtained.  The
balancing  degree desired is  achieved  by thc proper choice   of  the
standard element weights.  For example,  for $m  = 1$  we have  the
standard element weight distribution $1,2,3,...,  2,1$ and for  $m   =  2$
we get $1,3,4,4,..., 4,3,1$.  With increasing  $m$ the weight distribution
tends from a  trapezoidal  to a binomial  one, the latter corresponding to
$m = N - 1$.

\medskip

{\bf 4. Basic equation}

\medskip

      In  the  case  where the  object reaction  depends lin
early  on  the  incident  radiation  amplitude,  its respons
to a part of the flux  of the  ideal UR  (2) can  be writte
as follows:
\begin{equation}
J\Big(
\dac{1+k^2}{\ga^2}\Big) =\dil_0^{\iy}{\dac{d\Phi_{\De}(\eta)}{d\om'}
\Pi(\om)d\om'},
\end{equation}
where
\begin{center}
$\dac{d\Phi_{\De}}{d\om'}=\dac{d\Phi}{d\om'}\Big|_A -
\dac{d\Phi}{d\om'}\Big|_{\tilde{A}}, ~~d\om'=\dac{d\om}{\om}$.
\end{center}
$d\Phi/{d\om'}\big|_{A, \tilde{A}}$ - is the integral spectral density
of the photon flux for undulator magnetic field states  $A$ and
$\tilde{A}$, $\Pi (\om)$ is the spectral sensitivity of the object.

By the following change of variables in eq. (3):
\begin{center}
$\eta=e^\tau,
~~\om=\dac{2\pi c}{p}e^{-s}, ~~\dac{1+k^2}{\ga ^2}=e^x$,
\end{center}
we obtain the standard formula for the object reaction
to  the  UR  as  a convolu\-tion-type  Fredholm integral
eguation of the first kind:
\begin{equation}
U(x)=\dil_{-\iy}^{\iy}{K(x-s)Z(s)ds}.
\end{equation}
Here,
\begin{center}
$U(x)=J(e^x), ~~K(\tau)=\dac{d\Phi_{\De}}{d\om'}(e^{\tau}), ~~
Z(s)=\Pi\Big(\dac{2\pi c}{p}e^{-s}\Big)$.
\end{center}

Eq. (4) is a basic one in the MCS problem. In order
to solve this problem the object reaction is measured as
a function of the particle energy. The speetral density
$K(\tau)$ of the UR photon flux is measured experimentally
or calculated using known formulas. The spectral sensitivity of the object
$Z(s)$ is found from eq. (4) by numerical calculations using the
regularization methods which are applied to solve ill-defined problems.
   The angle-integral density of the UR photon flux
can be written down in the dipole approximation [5,7]
for $\si$- and $\pi$-polarization components:
\begin{center}
$\dac{d\Phi}{d\om'}\Big|^{\sigma}_{\pi}=\dac{4\pi L \alpha k^2}{3 T p}
\dac{I_1}{I_2}\phi(\eta)\Big|^{\sigma}_{\pi}$,

\medskip

$\phi(\eta)\Big|
^{\sigma}_{\pi}=\dac{4}{3I_1}\eta\dil_{\eta}^{\iy}{|H(\om_x)|^2 \left[
\begin{array}{ll}
1-2\dac{\eta}{\nu}+3\big(\dac{\eta}{\nu}\big)^2\\
1-2\dac{\eta}{\nu}+\big(\dac{\eta}{\nu}\big)^2\\
\end{array}
\right] \dac{d\nu}{\nu^2}},$
\end{center}
with the normalizing condition
\begin{center}
$\dil_{0}^{\iy}{\big[\phi(\eta)\big|_\sigma+\phi(\eta)\big|_\pi
\big]\dac{d\eta}{\eta}}=1.$
\end{center}
Here $\al\approx1/137$ is the fine
structure constant, $T$ is the orbit period for a charge in an accelerator
or storage ring,
\begin{center}
$I_j=\dil_{0}^{\iy}{\big|H(\om_x)\big|^2\dac{d\nu}{\nu^j}}, ~~
H(\om_x)=\dil_{-\iy}^{\iy}{H(x)e^{i\om_xx}dx}, ~~k^2=\Big(\dac{\mu e_0}{\pi
m_0 c}\Big)^2\dac{p}{L}I_2$,
\end{center}
is  the dipole  parameter, $e_0$ is the  electron charge,  $m_0$, is
the electron mass, $\om_x=\pi\nu/p, ~~\mu=4\pi\cdot10^{-7}~H/m$.  In  the
numerical  experiment  on  the  ATM   with  phase reversal and  weight
modulation  we   have  used   the  wide undulator approximation  which  is
close  to  that  in  practice, the  field being dependent only  on $x$.
The phase modulation  has been  realized by alternately switching of the
polarity of the undulator block supplies.  The criterion for the choice
of the correct modulation  parameters  $\{N, ~n, ~m, ~D/p\}$  is   the
vanishing integral difference given by:
\begin{equation}
\De=\dil_0^{\iy}{\dac{d\Phi_\De}{d\om'}(\eta)\dac{d\eta}{\eta}}=
\dac{4\pi\al}{3T}\Big(\dac{\mu e_0}{\pi m_0 c}\Big)^2
\dil_0^\iy{\big[|H(\om_x)|_A^2-|H(\om_x)|_{\bar{A}}^2\big]\dac{d\nu}{\nu}},
\end{equation}
where $n$ is the number of blocks and $D$ is the distance
between the neighbouring blocks. The geometric param-
eters of the standard elements and their number $M$ are
found from the basic initial parameters,  namely, the
undulator    length    and   the    undulator   gap.    Knowing   $M$
one  can  start  a  model  run  in  order  to  find  the  remaining
parameters   $\{N,  ~n,  ~m,  ~D/p\}$.   If  the   neighbouring  blocks
overlap, then $D < 0, ~D/p$ being an integer.
The  qualitative  criteria  for  the  results  of this  run are:
(i)  the  vanishing  $\De$  in  eq.  (5)  with  $D/p<0$;   (ii)  the
correctness  of  the  solution   of  the   MCS  problem   with  the
obtained UR kernel.

\newpage

{\bf 5. Conclusions}

\medskip

    (1)   When   the  motion   balancing  degree   $m$  increases,
an   effective  depression   of  the   low-frequency  component
of  the  UR kernel  $K(x -  s)$ in  the region  $x -  s <  0$ takes
place in the undulator blocks.

    (2)  When  $m$  increases,  localization  of  the   UR  kernel
in  the  region  $\eta\approx 1$  of the  high-frequency cut-off  of the
total   ideal  kernel   occurs.  This   monochromatization  effect  corresponds  to  the  angular  localization  of   the  UR
kernel in the angular range $\psi\in(0, ~1/\sqrt{N})$, in accordance
with the known formula $\nu=\eta(1+\psi^2)$.  This angular localization
near the direction $\psi=0$  causes the preferable selection of the
$\si$-component of UR polarization and  the strong  depression of  the
$\pi$-component, since  the intensity  of the  $\pi$-component is  small
for directions $\psi < 1\sqrt{N}$.

\medskip

{\bf  References}

\medskip

\begin{enumerate}
\item  E.E.  Koch  (ed.),  Handbook  on  Synchrotron  Radiation
(North-Holland, Amsterdam, 1983).
\item  M.M.  Nikitin  and  G.  Zimmerer,  Report DESY  SR 85-04
(Hamburg, 1985); Nucl. Instr. and Meth. A240 (1985) 188.
\item  V.G.   Bagrov,   A.F.   Medvedev,   M,M.  Nikitin   and  M.L.
Shinkeev, Nucl. Instr. and Meth. A261 (19B7) 337.
\item  A.F.  Medvedev,   M.M.  Nikitin   and  M.L.   Shinkeev,  Tomsk
Research Centre preprint N 2-90 (Tomsk, 1990) in Russian.
\item  M.M.   Nikitin   and    V.Ya.   Epp,    Undulator   Radiation
(Energoatomizdat, Moscow, 1988) in Russian.
\item  A.G.  Valentinov,   P.D.  Vobly   and  S.F.   Mikhailov,  INP
preprint N 89-174 (Novosibirsk, 1989) in Russian.
\item  L. Landau and E. Lifshitz, The Classical Theory of Fields
(Pergamon, London, 1962) section 77.
\end{enumerate}

\end{document}